\documentclass[12pt]{iopart}  % for review and submission
\usepackage{graphicx}  % needed for figures
\usepackage{dcolumn}   % needed for some tables
\usepackage{bm}        % for math
\usepackage{amssymb}   % for math

\newcommand{\fig}[1]{Fig.~\ref{#1}}

{}
\newcommand{\ww}{1.0}

\begin{document}

\title{Elastic Properties of Hybrid Graphene/Boron Nitride Monolayer}
\author{Qing Peng, Amir R. Zamiri, Wei Ji and Suvranu De}
\address{
Department of Mechanical, Aerospace and Nuclear Engineering,\\ Rensselaer Polytechnic Institute, Troy, NY 12180, U.S.A.\\
}

\begin{abstract}

Recently hybridized monolayers consisting of hexagonal boron nitride ($h$-BN) 
phases inside graphene layer have been synthesized 
and shown to be an effective way of opening band gap in graphene monolayers \cite{ISI:000276953500024}.
In this paper, we report a first-principles density functional theory study 
of the $h$-BN domain size effect on the elastic properties of 
graphene/boron nitride hybrid monolayers ($h$-BNC). 
We found that both in-plane stiffness and 
longitudinal sound velocity of $h$-BNC linearly decrease with $h$-BN concentration. 
Our results could be used for the design of 
future graphene-based nanodevices of the surface acoustic wave sensors and waveguides. 

\end{abstract}


\maketitle

\section{Introduction}

Substituting C with B and N atoms in graphene has been shown to be a promising way
to improve semiconducting properties of the graphene \cite{ISI:000265832400041,PhysReVLett.98.196803,PhysReVLett.101.036808}.
Hexagonal boron nitride ($h$-BN) monolayer \cite{Peng2012cms4} and graphene have similar 2D lattice structures but with very different physical properties. 
Interesting nanostructures can be made by mixing these two structures\cite{Kawaguchi:1197,Suenaga24101997,han:1110,ISI:000177977400012}. 

Recently, a promising method has been reported in which atomistic monolayers have been generated 
consisting of $h$-BN phases in graphene ($h$-BNC) using a thermal catalytic chemical vapor deposition method \cite{ISI:000276953500024}. 
These hybrid monolayers have been shown to have isotropic physical properties 
which can be tailored by controlling the kinetic factors 
affecting the $h$-BN domain size within graphene layer. 
This is different from B-doped or N-doped graphene, where the integrity of the $h$-BN structure is missing. 

The electronic band structures of the $h$-BNC heterostructures were studied in our previous work \cite{rpi1}. 
However, the mechanical properties of these heterogeneous nanosheets are still unknown. 
In this paper we investigate the elastic properties of the $h$-BNC monolayers 
as a function of $h$-BN concentration using {\em ab initio} density functional theory. 
%%The model and computational details are presented in Section II, followed by results and analysis in Section III and conclusions in Section IV.  

\section{Modeling and Computational Details}

The effect of $h$-BN domain on the properties of the $h$-BNC hybrid structures 
is modeled by only considering the eﬀect of its size  
while maintaining its hexagonal structure within the system.
The h-BN domain size effect
then can be represented by the h-BN concentration $x$ in the
model as (B$_3$N$_3$)$_x$(C$_6$)$_{1-x}$ where (B$_3$N$_3$) and (C$_6$)
denote the nanodomain structure of h-BN monolayer and graphene, respectively.
The proposed domain size effect model (B$_3$N$_3$)$_x$(C$_6$)$_{1-x}$ is based on
the results of previous studies of B$_x$C$_y$N$_z$ layered structures\cite{Mazzoni2006},
layers and nanotubes\cite{ISI:000286487300049,ISI:000290652200008},
quantum dots and nanorods\cite{ISI:000265030000048}, and monolayer nanohybrids\cite{ISI:000290914700066}.
It is a general belief that the $h$-BN segregates in the $h$-BNC, and the system gains 
lower energy, larger band gap, and better thermal stability after phase segratation \cite{ISI:000276953500024,ISI:000286487300049,ISI:000290914700066}.
The six-atom hexagonal structures (both B$_3$N$_3$ and C$_6$)
are the basic blocks in these heterostructures.
This model of (B$_3$N$_3$)$_x$(C$_6$)$_{1-x}$ captured the main feature of these heterostructures.
As shown by previous works \cite{Mazzoni2006,ISI:000286487300049}, 
both stoichiometry and geometry change the band gap of B-C-N nanotube. 
The maximum band gap is achieved at B/N ratio of 1. 
In the B-C-N monolayer, however, 
%both shape and size of the embedded nanodomain have significant effects in magnetic properties, 
the domain size is a dominant factor comparing with domain shape in changing the band gap \cite{ISI:000290914700066}.
By varying h-BN concentration $x$, the domain size effect
on mechanical properties of mono-layer hexagonal BNC heterostructures
can be appropriately
presented in our (B$_3$N$_3$)$_x$(C$_6$)$_{1-x}$ model.

We need to emphasis that although we only investigated the effect of stoichiometry of $h$-BN here,
this domain model is different from point model of B$_x$C$_y$N$_z$ model
at $x=z$ where hexagonal (B$_3$N$_3$) structure is not considered \cite{Mazzoni2006}.
In other words, our model specified not only the stoichiometry of B/N ratio of 1,
but also the hexagonal (B$_3$N$_3$) and C$_6$ structures,
to represent the two separated phases in heterogeneous h-BNC structures.

\begin{figure}
\includegraphics[width=\ww\textwidth]{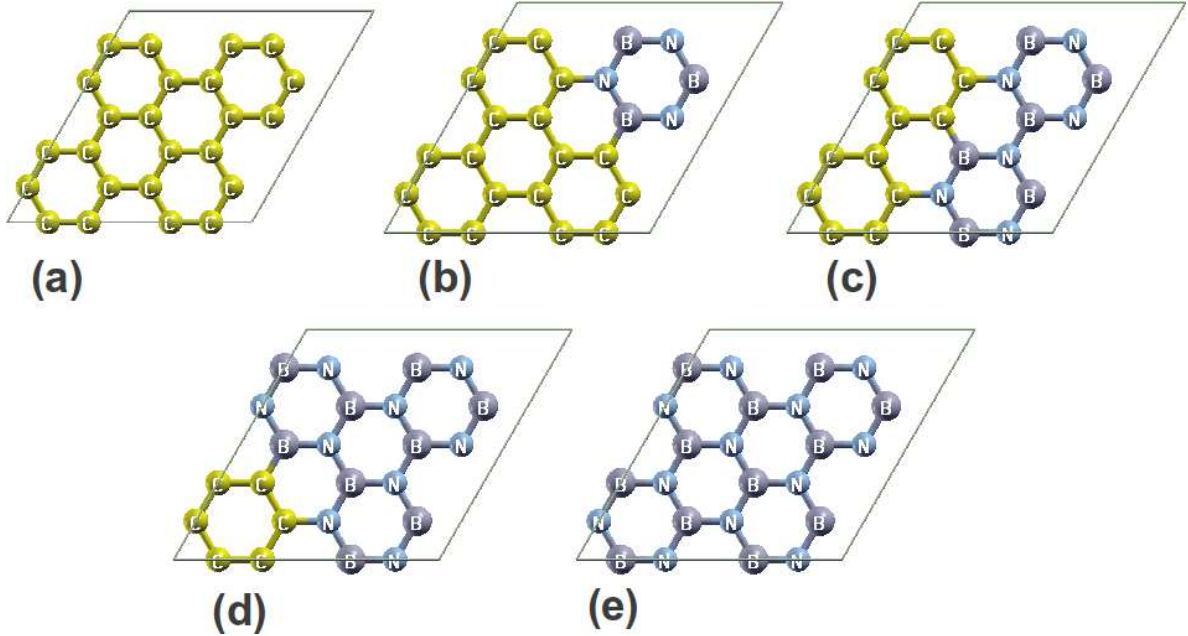}
\caption{\label{fig:config} Atomic structures of five configurations in order of $h$-BN concentration: (a) 0\%; (b) 25\%; (c) 50\%; (d) 75\%; (e) 100\%. } 
\end{figure}

We examined the change in elastic properties of the $h$-BNC monolayer as a function of $h$-BN concentration. 
Five $h$-BNC configurations, in order of $h$-BN concentration, 0\%, 25\%, 50\%, 75\% and 100\% have been studied, 
where 0\% and 100\% corresponding to pure graphene and $h$-BN, respectively. 
The three other concentrations were selected based on their simplicity and representativeness. 
The atomic structures of these five configurations (\fig{fig:config}) 
were determined by the {\em ab initio} density functional theory through geometry optimization. 

DFT calculations were carried out with the Vienna Ab-initio
Simulation Package (VASP)
\cite{VASPa,VASPb,VASPc,VASPd} which is based
on the Kohn-Sham Density Functional Theory (KS-DFT) \cite{DFTa,DFTb} 
with the generalized gradient approximations as parameterized by Perdew, Burke and Ernzerhof (PBE)  
for exchange-correlation functions \cite{GGA}.
The electrons that explicitly included in the calculations, are the ($2s^22p^2$) electrons of carbon, 
the ($2s^22p^1$) electrons of boron and the ($2s^22p^3$) electrons of nitrogen. 
The core electrons ($1s^2$) of carbon, boron and nitrogen were replaced by the projector augmented wave (PAW) and pseudopotential approach\cite{paw1,paw2}. 
A plane-wave cutoff of 520 eV was used in all the calculations, 
such value was chosen to eliminate the pulay stress during the geometry optimization \cite{Francis1990}.

The criterion for stopping the relaxation of the electronic degrees of freedom 
was set by total energy change to be smaller than 1.0$-6$ eV. 
The optimized atomic geometry was achieved through minimizing Hellmann-Feynman forces 
acting on each atom  until the maximum forces on the ions were smaller than 0.001 eV/\AA.

The atomic structures of the five configurations 
were obtained by fully relaxing a 24-atom-unit cell where all atoms were placed in one plane. 
The irreducible Brillouin Zone was sampled 
with a Gamma-centered $19 \times 19 \times 1$ $k$-mesh 
and initial charge densities were taken as a superposition of atomic charge densities. 
There was a 14 \AA \, thick vacuum region 
to reduce the inter-layer interaction to model the single layer system.

The elastic tensors were determined by performing six finite distortions of the lattice and 
deriving the elastic constants from the stress-strain relationships \cite{PhysRevB.65.104104}. 
In this study, we were interested only the in-plane elasticity. 
The strains of $\pm0.015$ were applied along two directions, $x$ and $xy$. 
Two stress-strain curves, 
$\sigma_{11}$ vs $\varepsilon_{11}$ and $\sigma_{12}$ vs $\varepsilon_{12}$, 
were computed, where  $\sigma_{11}$ and $\varepsilon_{11}$ 
are stress and strain in $xx$ direction, respectively, 
and $\sigma_{12}$ and $\varepsilon_{12}$ 
are the stress and strain in the $xy$ direction.
The elastic constants then were obtained through 
the least-squares extraction of coefficients from calculated stress-strain data. 
The final results of the values of these elastic constants were printed directly in VASP \cite{PhysRevB.65.104104}.

\section{Results and Analysis}

With the full geometry optimization, including allowing the change of shape, 
the final atomic structures of the five configurations 
were determined using DFT calculations, as shown in panels (a)-(e) in \fig{fig:config}.
The hexagonal structures were still kept, even for the configuration (c), which has 50\% $h$-BN. 
In addition, we tested another atomic structure of configuration (c), 
where the two gaphene rings are in diagonal, in stead of parallel along the side. 
The lattice constants of the two atomic structures are the same after geometry optimization, 
since the two atomic structures are symmetrical, 
and both have one layer of graphene ribbon in addition to one layer of $h$-BN ribbon, 
due to the periodic boundary conditions.

The lattice constants of the $h$-BNC mixtures were calculated as the half of the lattice vectors of the super-cells, 
for the convenience of comparison to those of the pure graphene and $h$-BN.
We found that the lattice constant increases with $h$-BN concentration, which is denoted as $C_{BN}$ in this paper.
Our results are in a good agreement with previous experimental results reported for $h$-BN (2.51 \AA) \cite{PhysRevB.68.104102}
and graphene (2.46 \AA) \cite{ISI:A1955WB70900014}.

\begin{figure}
\includegraphics[width=\ww\textwidth]{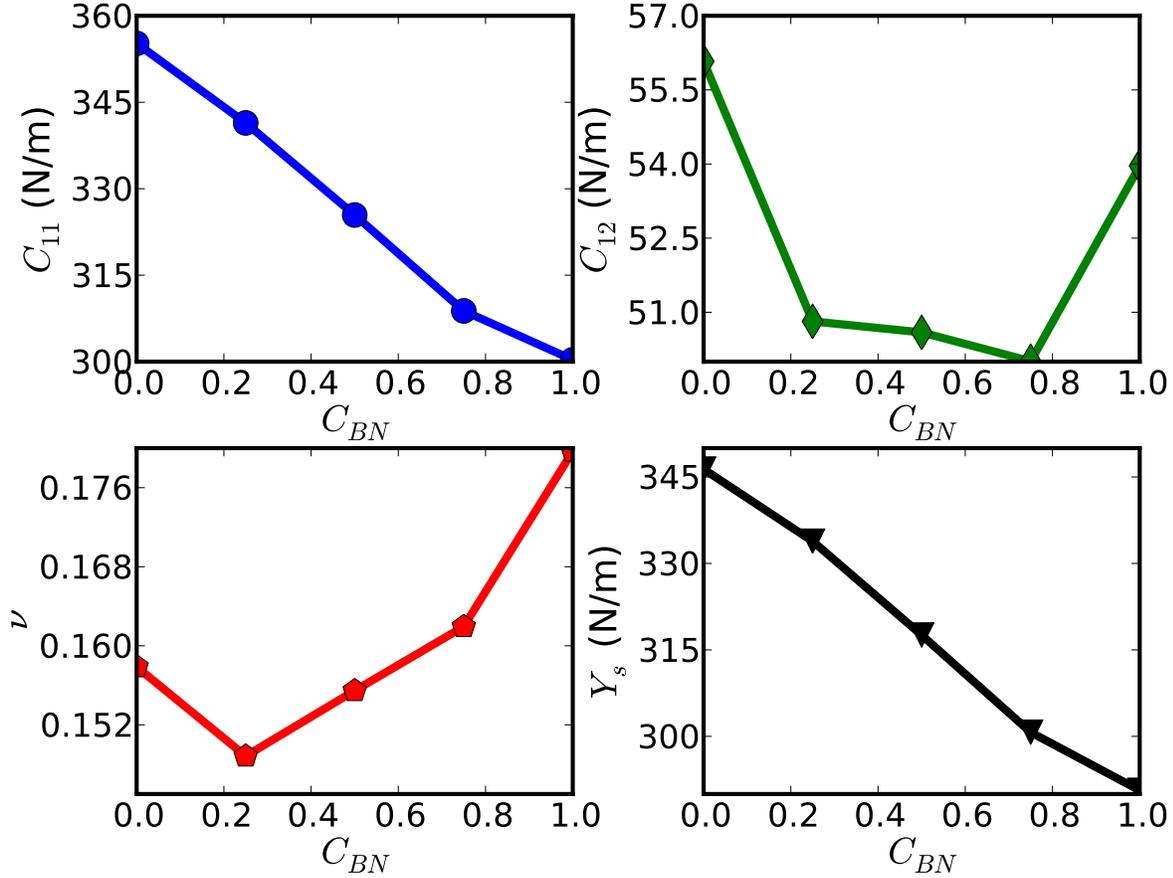}
\caption{\label{fig:elast} Elastic constants $C_{11}$ and $C_{12}$, Poisson's ratio $\nu$ and in-plane stiffness $Y_s$ as a function of $h$-BN concentrations $C_{BN}$. } 
\end{figure}

The elastic constants were obtained from DFT calculations. 
Due to the symmetry, only $C_{11}$ and $C_{12}$ are independent. 
$C_{11}$ decreases linearly with respect to $C_{BN}$, 
as plotted in top-left panel of \fig{fig:elast}. 
The in-plane stiffness $Y_s$
can be obtained from the elastic moduli $C_{11}$ and $C_{12}$ 
as $Y_s=(C_{11}^2-C_{12}^2)/C_{11}$.
The Poisson's ratio $\nu$ which is the ratio of the transverse strain 
to the axial strain can be obtained from elastic moduli as $\nu=C_{12}/C_{11}$.
Our result of $\nu$ and $C$ for the five configurations 
are shown in bottom panels of \fig{fig:elast}. 
Our calculated value for in-plane stiffness of graphene (347.2 N/m) 
is in a good agreement with the experimental value ($340\pm50$ N/m) \cite{ISI:000257713900044}, 
and theoretical predictions (348 N/m in ref.\cite{PRBwei2009} 
and 335 N/m in ref. \cite{ISI:000275246200020}). 
Our calculated value of $Y_s$ for $h$-BN (291.3 N/m) 
agrees with {\em ab initio} (GGA-PW91) prediction (267 N/m in ref. \cite{ISI:000275246200020}).
The calculated Poisson's ratio $\nu$ is 0.16 for graphene and 0.18 for $h$-BN, 
which agree with reported values of 0.16 and 0.21, respectively \cite{ISI:000275246200020}.

\begin{figure}
\includegraphics[width=\ww\textwidth]{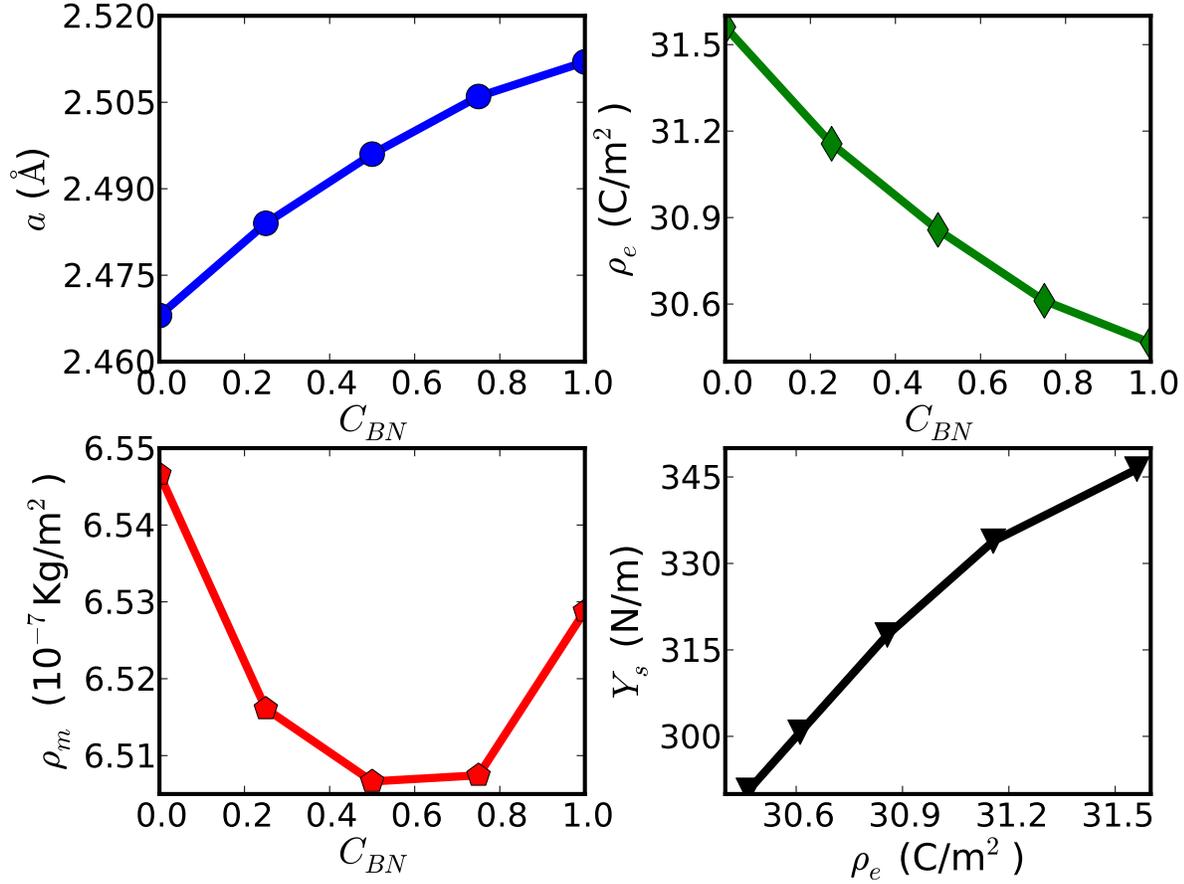}
\caption{\label{fig:rho} Lattice constants $a$, electronic charge density $\rho_e$ and mass density $\rho_m$ as a function of $h$-BN concentrations $C_{BN}$. The in-plane stiffness $Y_s$ varied with electronic charge density is plotted. } 
\end{figure}

As shown in \fig{fig:elast}, both the $C_{11}$ and the in-plane stiffness 
decrease nearly linearly as $h$-BN concentration increases. 
However, the $C_{12}$ and Poisson's ratio, show rather complicated behaviors.
The similar trend of the $C_{11}$ and the in-plane stiffness 
is due to the fact that $C_{11}$ is the dominant factor in computing the in-plane stiffness, 
about six times bigger than $C_{12}$. 
For the same reason,  Poisson's ratio $\nu$ has the similar trend as $C_{12}$. 
Therefore, our analysises only focus on the dominant factor $C_{11}$.

The mechanical behaviors are ultimately determined by the electronic structures. 
For a better understanding of the mechanism of linear elastic properties with respect to the $C_{BN}$, 
we studied the total electronic charge density $\rho_e=q/S$, 
where $q$ is the total electronic charge of the system,  
and $S$ the area (cross section) of the system in the basal plane. 
We found that $\rho_e$ monotonically decreases with $C_{BN}$, 
as plotted in the top-right panel in \fig{fig:rho}. 
This relationship is consistent with the $a$-$C_{BN}$, 
since $q$ is the same in all five configurations. 
The area mass density $\rho_m=m/S$ ($m$ is the atomic mass) are also studied, 
as plotted in the bottom-left panel in \fig{fig:rho}. 
$\rho_m$ reaches the minimum at $C_{BN}=0.5$. 

The relationship between $Y_s$ and $\rho_e$ was examined and plotted in bottom-right panel of \fig{fig:rho}. 
It is interesting to note that the in-plane stiffness monotonically 
increase with respect to the area charge density $\rho_e$. 
It gives a hint that the elastic properties can be engineered 
by doping or introducing defects which changes the charge densities.

In these $h$-BNC structures, there is a non-zero stiffness both for volumetric and shear deformations. Hence, 
it is possible to generate sound waves with different velocities dependent on the deformation mode. 
Sound waves generating volumetric deformations (compressions) and shear deformations 
are called longitudinal waves ($p-$wave) and shear waves ($s-$wave), respectively.  
The sound velocities of these two type waves are respectively given by:
\cite{acoustics} 
\begin{equation}
v_p=\sqrt{\frac{Y_s(1-\nu)}{\rho_m(1+\nu)(1-2\nu)}}
\end{equation}
\begin{equation}
v_s=\sqrt{\frac{C_{12}}{\rho_m}}
\end{equation}

\begin{figure}
\includegraphics[width=\ww\textwidth]{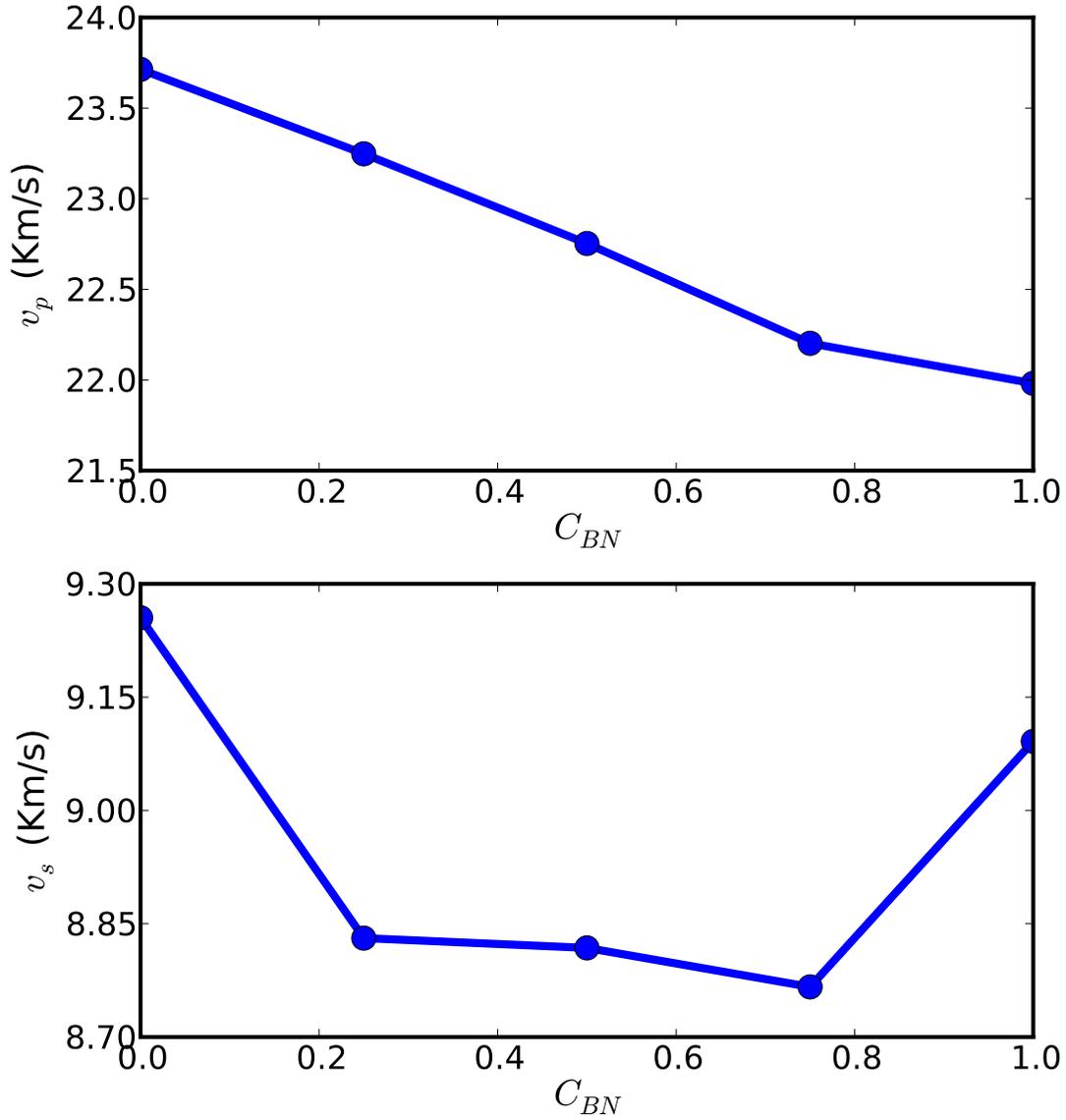}
\caption{\label{fig:wv} $p-$wave and $s-$wave velocity as a function of $h$-BN concentrations $C_{BN}$. } 
\end{figure}

The dependence of $v_p$ and $v_s$ on $C_{BN}$ are plotted in \fig{fig:wv}. 
$v_p$ monotonically reduced with $C_{BN}$.
However, $v_s$ has a more complex behavior. 
The $v_s$ of $h$-BNC decreases with $C_{BN}$, 
but is lower than that of the pure phases, both graphene and $h$-BN.
The $v_p/v_s$ is in the range between 2.42 and 2.64.

Since the speed of these sound waves can be measured experimentally when these $h$-BNC heterostructures are synthesized, 
the predicted $p-$wave and $s-$wave velocities could be served as the quantities for the validation of the elastic properties. 

Furthermore, we predicted a linear relationship between the sound velocity and the concentration of $h$-BN. 
As a result, a sound velocity gradient can be achieved by introducing $h$-BN domains into graphene. 
The sound velocity gradient could be used to form a sound frequency and ranging channel, 
which is the functional mechanisms of surface acoustic wave (SAW) sensors and waveguides.
Thus, graphene-based nano-devices of SAW sensors and waveguides can by synthesised with $h$-BNC for next generation electronics.  

\section{Conclusion}

  In summary, we used {\em ab initio} density functional theory 
to investigate the effect of the $h$-BN domain size on the 
elastic properties of graphene/$h$-BN hybrid monolayers. 
The elastic constants of five configurations were explicitly examined. 
We found that the in-plane stiffness increases linearly 
with the $h$-BN concentrations. 
We predicted a linear relationship between the sound velocity and the concentration of $h$-BN. 
This knowledge could be used for the design of 
future graphene-based nanodevices of the surface acoustic wave sensors and waveguides.
Thus our results may provide guidance in practical
engineering applications of these nano-heterostructures.

\ack
The authors would like to acknowledge the generous financial support from the Defense Threat Reduction Agency (DTRA) Grant \# BRBAA08-C-2-0130.

\section*{References}
\bibliography{elast}

\end{document}